\documentstyle[12pt]{article}
\newskip\humongous \humongous=0pt plus 1000pt minus 1000pt
\def\caja{\mathsurround=0pt}

\def\eqalign#1{\,\vcenter{\openup1\jot \caja
        \ialign{\strut \hfil$\displaystyle{##}$&$
        \displaystyle{{}##}$\hfil\crcr#1\crcr}}\,}
\newif\ifdtup

\def\be{\begin{equation}}
\def\ee{\end{equation}}
\def\ba{\begin{eqnarray}}
\def\ea{\end{eqnarray}}

\begin{document}
\renewcommand{\theequation}{\thesection.\arabic{equation}}
\newcommand{\beq}{\begin{equation}}
\newcommand{\eeq}[1]{\label{#1}\end{equation}}
\newcommand{\ber}{\begin{eqnarray}}
\newcommand{\eer}[1]{\label{#1}\end{eqnarray}}
\begin{titlepage}
\begin{center}

\hfill NSF-ITP-95-129\\
\hfill CPTH-S379.1095 \\
\hfill hep-th/9510094\\

\vskip .5in

{\large \bf  NATURAL GAUGE-COUPLING
UNIFICATION AT THE STRING SCALE}
\vskip .5in

{\bf C. Bachas $\ ^{a,b}$, C. Fabre $\ ^{b}$
and T. Yanagida $\ ^{a ,c}$}
\vskip .1in

{$\ ^{a}$
\em Institute for Theoretical Physics,
University of California, Santa Barbara, CA 93106, USA \\
$\ ^{b}$
 Centre de Physique Th\'eorique, Ecole Polytechnique,
 91128 Palaiseau, FRANCE \\
$\ ^{c}$
  Department of Physics, University of Tokyo, Tokyo 113, JAPAN
}

\vskip .15in

\end{center}

\vskip .4in

\begin{center} {\bf ABSTRACT }
\end{center}

\vskip .15in
We argue that  moduli in the adjoint
 representation of the standard-model gauge group
are a natural feature of superstring models,
and that they can account for the apparent
discrepancy between the string and unification scales.

\begin{quotation}\noindent
\end{quotation}
\vskip1.0cm
NSF-ITP-95-129 \\
CPTH-S379.1095\\
October 1995\\
\end{titlepage}
\vfill
\eject
\def\baselinestretch{1.2}
\baselineskip 16 pt
\noindent
 \setcounter{equation}{0}

An important hint in favour of supersymmetric GUTs is the
unification of gauge couplings extrapolated from their
low-energy experimental values under the assumptions of
minimal particle content and superpartner thresholds at
$\sim 1 TeV$ \cite{Dimo}.
The scale at which the couplings meet ($M_{GUT}^{minimal} \sim 2\times
10^{16} GeV$)
 lies,  however, one  order of magnitude below
the heterotic string scale  ($M_{str} \sim 0.5\ g_{str}
\times 10^{18} GeV$) \cite{Kapl},
  suggesting that there should exist some string states (beyond those
of the MSSM) which are
significantly lighter than $M_{str}$
\footnote{Note however that their mass can  be pushed up if
they are very numerous \cite{Nill}. They
may even be unnecessary if one allows a non-standard hypercharge
normalization and an $\alpha_3$ at the higher end of its
allowed range \cite{Iban}.
}.
It has been proposed that these
light states can be exotic vector-like quarks and leptons
with non-conventionnal assignment of hypercharge
\cite{Dien},
or else the extra gauge bosons and Higgses of a unifying symmetry
broken somewhere below $M_{str}$
\cite{Lewe,Barb}.
In this note we would like to point out a natural alternative:
  a color-$SU(3)$ octet and a color-neutral triplet of  weak-$SU(2)$,
both having
zero hypercharge.
As we will argue {\it (a)} these
 appear  in many string models as continuous moduli
which is why they can remain  light naturally, and
{\it (b)} they
 push the unification scale up to $M_{str}$ when their masses
lie in the best-motivated intermediate range
$M_{Plank}^{2/3} m_{susy}^{1/3} \sim 10^{13} GeV$.
Furthermore contrary to exotic stable remnants
they present no danger for cosmology.

The one-loop running coupling constants in the presence of
an adjoint scalar multiplet  read \cite{Yana}:
$$
\eqalign{
{2\pi\over\alpha_1(\mu)} - {2\pi\over\alpha_1(m_Z)} &=
({1\over 2}+{2\over 3}N_g) log({m_{susy}\over m_Z}) - (2N_g + {3\over 5})
log({\mu\over m_Z})
\cr
{2\pi\over\alpha_2(\mu)} - {2\pi\over\alpha_2(m_Z)} &=
({4\over 3}+{2\over 3}N_g +{5\over 6}) log({m_{susy}\over m_Z})+
\cr
& \ \ \ \ \ \ \ \ \ \ \ \ \
\ \ \  + (6-2N_g - 1)
log({\mu\over m_Z}) - 2 log({\mu\over m_{(3)}})
\cr}
$$
and
$$\eqalign{
{2\pi\over\alpha_3(\mu)} - {2\pi\over\alpha_3(m_Z)} &=
(2 +{2\over 3}N_g )log({m_{susy}\over m_Z}) + \cr
& \ \ \ \ \ \ \ \ \ \ \ \ \ \  \ \   +(9-2N_g)
log({\mu\over m_Z})- 3 log({\mu\over m_{(8)}})
\cr}
$$
where $m_{(8)}$ and $m_{(3)}$ are the masses of the color octet
and weak triplet, respectively.
Using as input the experimental values of the couplings at $m_Z$
(with  $\alpha_3 = 0.116\pm 0.005$) \cite{Erle}
and assuming as usual $m_{SUSY} \simeq 1 \ TeV$,
one can calculate from the above equations the required masses
of these extra particles
such that couplings  unify at the string scale.
Taking $ M_{str} \simeq 6\times 10^{17} GeV$,  corresponding
to a $k=2$ Kac-Moody level,
\footnote{Adjoint scalars require that $k$ be at least equal to two.
Higher values push the string scale up
and would demand somewhat lighter intermediate masses.
}
we find:
$m_{(3)}\simeq 5.3\times 10^{12} GeV$ and
$m_{(8)}\simeq 2.8\times 10^{12} GeV$
{}.
Surprisingly enough these masses are
not only close to each other,
but also of the   order of magnitude
one  would expect if supersymmetry breaking were
induced by  condensing gauginos
\cite{Dere}. Our underlying assumptions are of course {\it (i)}
standard hypercharge normalization as in the usual SU(5) embedding,
and {\it (ii)} zero vevs for the adjoint moduli, consistent
with an unbroken standard model gauge group.

Such adjoint moduli are an ingredient of all the
recently constructed models based on a group structure
$G\times G$  \cite{Masl,Alda,Finn}.
They are the relics after truncation  of extended
supersymmetric vector multiplets, which is why their
 potential stays flat \cite{ABK}.
To illustrate this point explicitly consider a $N=2$
supersymmetric  pure gauge theory with  gauge group $SO(2n)$.
We may define in this theory  two
 parity operations,  $(-)^A$  and ${\cal P}$,
 where $A$
counts the number of vector indices under the first factor
in the embedding $SO(2n) \supset SO(n)_A \times SO(n)_B$,
while
  ${\cal P}$  breaks down a $N=2$ vector multiplet into
an even vector and an odd scalar superfield of $N=1$.
Imposing  the combined parity
projection $(-)^A {\cal P} = +1$ leads to
a truncated $N=1$ supersymmetric theory with
 gauge group  $SO(n)_A \times SO(n)_B$
and with extra chiral multiplets in the representation $(n,n)$.
A heritage of the extended supersymmetry
is that these latter have a flat potential along directions
lying inside the Cartan subalgebra of the original $SO(2n)$.
By turning on a non-vanishing vev for the trace of the
$(n,n)$ matrix,
we can break the gauge symmetry to the  diagonal subgroup
$SO(n)_{diag}$.
 As a minute's thought
will convince the reader, the resulting theory still has
$SO(n)_{diag}$-adjoint moduli.
 Note that the level of the
corresponding algebra
is the sum of the levels of the two group factors, and hence
it is at least equal to two. Note also that in
  string theory a  $Z_2$ truncation  gives   rise to
new (twisted) states.  These however only appear in pairs, and
thus do not spoil the flat directions as long as their
expectation values vanish.

  The original motivation \cite{Lewe,Barb} for constructing $G\times G$
string models was to allow minimal unification by enlarging
the symmetry at a scale significantly lower than $M_{str}$.
 As we have just argued the existence of adjoint moduli turns
this motivation around on its head: it invalidates the desert
hypothesis and, in the absence of other light states,  can render
 premature unification
 unnecessary.
Our argument also carries over
 to those string GUT models,
constructed directly at $k>1$ \cite{Alda,Chau}, which
are just stringy realizations of the  last
$G\times G\rightarrow H_{diag}$ breaking step \cite{Lewe}.
More generally,
  it is hard to imagine how a unifying
gauge symmetry can break at a scale an order of magnitude
below  $M_{str}$
if not along some flat directions which would leave
behind light extra matter  populating  the desert.
Thus it might be better motivated
(and safer with respect to proton decay) to search
 for models whose observable light states include
  only a $k>1$ MSSM and adjoint moduli.

\vskip 1cm
\noindent
{\bf Acknowledgments} \\
This work was supported by the NSF under grant no. PHY94-07194, and
by EEC grants CHRX-CT93-0340 and SC1-CT92-0792. We thank
our colleagues at the ITP workshop for
enlightening discussions, and in particular
 K. Dienes and P. Langacker
for a careful reading of the manuscript.

\end{document}